\documentclass{article}

\begin{document}

\title{Photon-Axion-Like Particle Coupling Constant and
Cosmological Observations.}

\author{Piotrovich M. Yu., Gnedin Yu. N.\thanks{E-mail: gnedin@gao.spb.ru},
Natsvlishvili T. M.\\ Central Astronomical Observatory at Pulkovo,
Saint-Petersburg, Russia.}

\maketitle

\begin{abstract}
We estimated the photon-pseudoscalar particle mixing constant from
the effect of cosmological alignment and cosmological rotation of
polarization plane of distant QSOs, revealed by Hutsemekers et al.
\cite{r1}. This effect is explained in terms of birefringent
phenomenon due to photon-pseudoscalar (axion-like) particle mixing
in a cosmic magnetic field.

On the contrary, one can estimate the strength of the cosmic
magnetic field using the constraints on the photon-axion-like
particle coupling constant from the CAST experiment and from SNe
Ia dimming effect \cite{r2}. In a result, the lower limit on the
intergalactic ($z\approx 1\div 2$) magnetic field appears at the
level of about $4\times 10^{-10}\div 10^{-11}$ G.

\smallskip

\end{abstract}

\section{Introduction}

One of the most intrigue recent astronomical discoveries was
detecting extreme-scale alignments of quasar polarization vectors
by Hutsemekers et al. \cite{r1,r3}. Based on sample of 355 quasars
with significant optical polarization and using various
statistical methods, they revealed that quasar polarization
vectors are not randomly oriented over the sky as it could
expected. The probability of this effect is often in excess of
99.9\%. The polarization vectors appear coherently oriented over
huge ($\sim 1$Gpc) cosmological regions of the sky located at both
low ($z\sim 0.5$) and high ($z\sim 1.5$) redshifts and
characterized by strongly different preferred directions of quasar
polarization. Two possible interpretations of this effect have
been proposed. The first one is a weakly developed interpretation
like a global rotation. But the properties that Hutsemekers et al.
\cite{r1} observed correspond qualitatively better to the
dichroism and birefringence produced by photon-pseudoscalar
oscillation within the intergalactic magnetic field \cite{r4}.

In order to further study the reality of this alignment effect,
Hutsemekers et al. \cite{r1} have subsequently carried out another
test which consisted in obtaining new polarimetric measurements
located in a region of the sky where the first preliminary result
of the alignment were revealed. These new measurements
independently confirmed the existence of coherent orientations of
quasar polarization vectors in the considered regions of sky.
Another exciting result obtained by Hutsemekers et al. \cite{r1}
is that the alignment effect seems to be close to an axis not far
from preferred directions tentatively identified in the CMB maps.
These conclusions have been made at the base as of new
polarization observation of a new sample of 335 polarized quasars
with accurate linear polarization measurements, so of combined
recent data from the literature. The most part of the polarimetric
observations were carried out at the European Southern Observatory
(Chile).

Although the interpretation of the alignment effect remains
uncertain, the effect itself appear as new way to test the
Universe and dark energy and dark matter components at large
scale. This effect presents also possibility to investigate the
physical properties of Intergalactic Medium (IGM). The last
interesting result has been recently presented by Borguet et al.
\cite{r5}. They found that the orientation of the rest-frame
UV/blue extended emission is correlated to the direction of the
quasar polarization and this effect may be connected with Type
1/Type 2 dichotomy of QSO host galaxies.

The final conclusion made by Hutsemekers et al. \cite{r1} is that
quasar polarization angles are definitely not random oriented over
the sky. Polarization vectors appear coherently oriented over very
large spatial scales in regions located at both low and high
redshifts and characterized by different preferred directions.
Authors claimed that polarization vector alignment seems connected
on a sizeable fraction of the known Universe points towards a
global mechanism acting at the scale of the Universe. They noted
also that the observed behaviour quite well corresponds to the
dichroism and birefringence predicted by photon-pseudoscalar
oscillation within an intergalactic magnetic field. If it is right
then we might have found a signature of either dark matter or dark
energy.

One should mention that the linear dichroism of aligned
interstellar dust grains in our Galaxy produces also linear
polarization along the line of sight. This polarization can
contaminate to some extent the quasar measured data and this may
change their position angles. However Sluse et al. \cite{r6}
investigated carefully this effect and showed that interstellar
polarization has a little effect on the polarization angle
distribution of significantly polarized ($p_l\geq 0.6\%$) quasars.

From the point of view photon-pseudoscalar (ultra light axion or
axion-like pseudoscalar particle) mixing in a magnetic field seems
as a quite promising interpretation (see \cite{r7,r8,r9,r10,r4})
especially because many of the observed properties of the
alignment effect were qualitatively predicted. The last PVLAS
Collaboration experiment reported the upper limit of observation
of a laser light polarization rotation in vacuum in the presence
of a transverse magnetic field is increasing the interest to this
phenomenon and to search of its astrophysical manifestations
(\cite{r11,r4}).

Here we estimate the value of the constant coupling of
photon-pseudoscalar mixing from the data of the cosmological
rotation of polarization plane of the intrinsic polarization of
QSOs in the intergalactic magnetic field (IGMF).

\section{Photon-pseudoscalar mixing and birefringent effect\\ in a magnetic field: the basic equations}

The probability of the magnetic conversion of photons into
low-mass pseudoscalar (scalar) particles were calculated by
Raffelt and Stodolsky \cite{r12} (see also \cite{r13}). The case
of massless pseudoscalars (arions) was developed by Anselm and
Uraltsev \cite{r14} (see also \cite{r15,r8}).

The expression for this probability takes a form:

\begin{equation}
P(\gamma_{\|}\leftrightarrow
a)=\frac{1}{1+x^2}\sin^2{\left(\frac{1}{2}B_{\bot}g_{a\gamma}L\sqrt{1+x^2}\right)}
\label{eq1}
\end{equation}

\noindent where

\begin{equation}
x=\frac{(\varepsilon -1)\omega}{2B_{\bot}g_{a\gamma}}
\label{eq2}
\end{equation}

\noindent $\omega$ is the radiation frequency, $\varepsilon$ is
the dielectric function of a medium that the light is propagating
through, $B_{\bot}$ is the magnetic field component perpendicular
to the photon direction, $g_{a\gamma}$ is the coupling constant
between photons and pseudoscalars.

The essential features of this probability functions are: (a) its
oscillatory character, and (b) the fact that the conversion
process is very sensitive to the polarization state of photon. The
last fact means that only a single polarization state with the
electric vector oscillating into the plane of directions of the
magnetic field and the photon propagation is subject to the
conversion.

The conversion process depends strongly on the dielectric function
of a medium. Assuming that the medium is a plasma with an electron
density $N_e$ and a neutral gas (for example, hydrogen) with
density $N_H$, we find:

\begin{equation}
1-\varepsilon = \frac{\omega^2_p}{\omega^2} - 4\pi\beta N_H -
\frac{28\alpha^2}{45m^2_e}B^2 - \frac{m^2_a}{\omega^2}
\label{eq3}
\end{equation}

The (\ref{eq3}) presents the most common expression for the
dielectric function, including the plasma polarizability (the
first term) and the contribution of a neutral gas ($\beta$ is the
atomic polarizability). The third term describes the contribution
of polarizability of a vacuum in a magnetic field. The last term
takes into account the contribution of the scalar field describing
the pseudoscalar particle. For the massless bosons (arions) it
equals to zero.

In Eqs.(\ref{eq1}-\ref{eq3}) we use the Loretz-Heaviside system of
units with $\hbar = c = 1$ and a fine structure constant $\alpha =
e^2/4\pi = 1/137$. In this system of units one gauss corresponds
to $6.9\times 10^{-2}(eV)^2$ and 1 cm corresponds to $5\times 10^4
(eV)^{-1}$.

Photon-pseudoscalar mixing also yields birefringent effect in a
magnetic field because the change of parallel polarization mode is
produced via the conversion process of photons into pseudoscalars.
Therefore the plane of polarization will be oscillated and
ellipticity will be acquired by a linear polarized beam
propagating across magnetic field lines in vacuum. In a result one
can present the following expression for rotation angle $\theta$
for a homogeneous case:

\begin{equation}
\tan{\theta(L)} =
\frac{1}{2(1+x^2)}\sin^2{\left(\frac{1}{2}B_{\bot}g_{a\gamma}L\sqrt{1+x^2}\right)}
\label{eq4}
\end{equation}

For a pure vacuum and the weak mixing the (\ref{eq4}) transforms
to:

\begin{equation}
\theta (L) = \frac{1}{8}g_{a\gamma}^2B_{\bot}^2L^2
\label{eq5}
\end{equation}

Later on, we are using (\ref{eq5}), but are taking into account
the dependence of the basic physical parameters on the
cosmological redshift $z$.

For a cosmologically distant source the change in polarization
angle $\Delta\theta$ due to propagation through intergalactic
medium with a magnetic field can be presented as:

\begin{equation}
\Delta\theta(z)\approx
\frac{1}{8}\left(g_{a\gamma}\int^{z_1}_{z_0}B_{\bot}(z)\frac{dL}{dz}dz\right)^2
\label{eq6}
\end{equation}

\begin{equation}
L(z) = \frac{c}{H(z)} =
\frac{c}{H_0}\int^z_0\frac{d\acute{z}}{(\Omega_m(1+\acute{z})^3+\Omega_{\Lambda})^{1/2}}
\label{eq7}
\end{equation}

We use the standard cosmological model: $\Omega_m = 0.27$,
$\Omega_{\Lambda}=0.73$ and $H_0 = 75 km/s Mpc$. We consider the
situation when the light beam is propagating in the IGM
(InterGalactic Medium) through the cosmological distance $\Delta z
= z_1 - z_0$. In the special case, one can consider the situation
when $z_0 = 0$. But here we shall consider the situation presented
by Hutsemekers et al. \cite{r1}, namely, $z_0 = 0.5$, $z_1 = 1.5$
and $z_0 = 1$, $z_1 = 2$. The typical value of the rotation angle
obtained by Hutsemekers et al. \cite{r1} from the polarimetric
observations of quasars was to be $\Delta\theta\approx 0.5$
($\Delta\theta\approx 30^{\circ}$) for the examined redshifts
$z_0$ and $z_1$.

Let us remind also that we consider the polarization rotation
effect due to the process of magnetic conversion of a photon into
a pseudoscalar particle. In our case unlike on the Faraday
rotation effect there is no strong dependence of a rotation angle
on the light wavelength. The absence of such kind dependence means
that the oscillation length via the photon conversion process is
less compare to other physical oscillation lengths and corresponds
to characteristic scale of the intergalactic magnetic field. The
numerical estimates will be presented in the next sections.

\section{Magnetic photon conversion in the IGM}

The Eqs.(\ref{eq5}-\ref{eq7}) are acting if the coherence length
$L_B$ in the IGM is determined by the magnetic field strength. It
means that the coherence lengths connected with plasma oscillation
frequency and pseudoscalar mass must be essentially higher than
$L_B$. It derives the following relations:

\begin{equation}
L_B = \frac{\pi}{g_{a\gamma}B}\ll L_p =
\frac{\pi\omega}{\omega^2_p},\,\,\, L_B\ll L_m =
\frac{\pi\omega}{m^2_a}
\label{eq8}
\end{equation}

\noindent where $\omega_p$ is the plasma frequency in IGM and
$m_a$ is the mass of a pseudoscalar (axion-like) particle.

The basic problem for the solution of (\ref{eq6}) is to find the
dependence of the intergalactic magnetic field on the cosmological
redshift, i.e. $B_{\bot}(z)$.

The first model, that will be tested, is the dependence
corresponding to equipartition between magnetic and CMB radiation
pressures:

\begin{equation}
B_{\bot}(z) = B_0 (1+z)^2
\label{eq9}
\end{equation}

It should be mentioned that such kind dependence have been
determined by Gnedin and Silant'ev \cite{r16} for quasar magnetic
fields from the observed decrease in the fraction of polarized
quasars with increasing redshift. Their conclusion has been given
at the base of observations by Impey et al. \cite{r17} and Wills
et al. \cite{r18}.

For the redshift dependence law of (\ref{eq9}) we obtain the
following estimate of the rotation angle for QSOs lying in the
redshift interval of $0.5\leq z \leq 1.5$:

\begin{equation}
\sqrt{\Delta\theta} =
0.244\frac{cB_0g_{a\gamma}}{H_0\sqrt{\Omega_m}} =
\left(\frac{g_{a\gamma}}{6\times 10^{-23}(eV)^{-1}}\right)
\left(\frac{B_0}{10^{-9}G}\right)
\label{eq10}
\end{equation}

For the redshift interval $1.0\leq z \leq 2.0$ we obtain

\begin{equation}
\sqrt{\Delta\theta} =
0.257\frac{cB_0g_{a\gamma}}{H_0\sqrt{\Omega_m}} =
\left(\frac{g_{a\gamma}}{4\times 10^{-23}(eV)^{-1}}\right)
\left(\frac{B_0}{10^{-9}G}\right)
\label{eq11}
\end{equation}

The next redshift dependence law that will be tested is,
so-called, the Chandrasekhar-Fermi law (\cite{r19}). Their law is
based on an interpretation of the dispersion in the observed
planes of polarization of the light of the distant objects.

Chandrasekhar and Fermi \cite{r19} have suggested the following
relation for the magnetic field component projected on the plane
of sky perpendicular to the line of sight:

\begin{equation}
B_{\bot} = \sqrt{\frac{4\pi}{3}\rho} \frac{V_{turb}}{\Delta\theta}
\label{eq12}
\end{equation}

\noindent and instead of $V_{turb}$ one uses usually the value of
dispersion of observed velocities, i.e. $V_{turb} = \Delta
V_{FWHM}$.

For the density of intergalactic medium we use the value of the
baryon density of the Universe: $\rho = n \Omega_b \rho_{cr}$,
where the cosmological parameter $\Omega_b = 0.04$ and the
numerical coefficient $n\sim 1$. In a result, the relation
(\ref{eq12}) transforms to

\begin{equation}
B_{\bot} = 1.2\times 10^{-15} (1+z)^{3/2}
\frac{V_{FWHM}}{\Delta\theta} \label{eq13}
\end{equation}

We use the ratio $\Delta V_{FWHM} = (\sqrt{3}/2)V_{FWHM}$.

To calculate the Eq.(\ref{eq6}), we accept the next values of IGM
temperature: $T_{IGM} = 10^5$K. This value is characteristic for
warm-hot intergalactic gas and has been obtained from estimates of
the ionizing spectral energy distribution as of nearby active
galactic nuclei so high-redshift quasars which demonstrate
Gunn-Peterson effect (\cite{r20,r21}). With use the relation
$V_{FWHM}\approx V_{thermal}$, we are obtaining from (\ref{eq6})
the following expression for the rotation angle ($0.5\leq z\leq
1.5$):

\begin{equation}
(\Delta\theta)^3 = 10^{-2}\left(\frac{g_{a\gamma}}{10^{-11}
GeV^{-1}}\right)^2 \left(\frac{B_0}{10^{-9}G}\right)^2
\left(\frac{75}{H_0}\right)^2 \left(\frac{T_{IGM}}{10^5 K}\right)
\label{eq14}
\end{equation}

Using the rotation angle value $\Delta\theta\approx 0.5$ for $z_0
= 0.5$ and $z_1 = 1.5$ from the data of Hutsemekers et al.
\cite{r1}, we obtain the next constraints for the coupling
constant $g_{a\gamma}$:

\begin{equation}
g_{a\gamma} \leq 3.4\times 10^{-11} GeV^{-1}
\left(\frac{10^{-9}G}{B_0}\right)\left(\frac{H_0}{75}\right)
\left(\frac{10^5 K}{T_{IGM}}\right)
\label{eq15}
\end{equation}

For intergalactic magnetic field with redshift dependence of
(\ref{eq9}) we have:

\begin{equation}
g_{a\gamma} \leq 4\times 10^{-14} GeV^{-1}
\left(\frac{10^{-9}G}{B_0}\right)\left(\frac{H_0}{75}\right)
\label{eq16}
\end{equation}

\section{Constraints on axion-like particle fundamental parameters}

It is evident that in our situation there exists the strong
dependence of obtained constraints on the redshift dependence of
the intergalactic magnetic field. In a result we obtain the
''soft'' upper limit $g_{a\gamma}\leq 3.4\times 10^{-11} GeV^{-1}$
and the ''hard'' upper limit $g_{a\gamma}\leq 4\times 10^{-14}
GeV^{-1}$ of the coupling constant between axion and photon
fields.

The next important constraint can be obtained for the mass of a
boson (axion-like particle). This constraint can be derived from
Eq.(\ref{eq8}). This equation requires that the magnetic
pseudoscalar-photon mixing coherence length $L_B$ will be less
that the coherence length due to the pseudoscalar mass $L_m \sim
m^{-2}_a$. In a result, we obtain the constraints on the mass
magnitude of a pseudoscalar.

The Eqs.(\ref{eq8}), (\ref{eq15}) and (\ref{eq16}) derives:

\begin{equation}
m^2_a \ll \omega g_{a\gamma} B
\label{eq17}
\end{equation}

\noindent and

\begin{equation}
m_a \ll 2.5\times 10^{-15} eV\,\,\, for\,\,\, B_{\bot} = B_0
(1+z)^{3/2} \label{eq18}
\end{equation}

\begin{equation}
m_a \ll 8\times 10^{-16} eV\,\,\, for\,\,\, B_{\bot} = B_0 (1+z)^2
\label{eq19}
\end{equation}

These upper bounds are more stringent than bounds obtained from
the absence of high energy gamma-rays from SN 1987A (see, for
example, \cite{r22}) and from the CAST experiment (\cite{r23}).
Our results agree better with the propose that the observed
faintness of high redshift supernovae could be attributed to the
mixing of photons with a light pseudoscalar (light axion-like)
particles in an intergalactic $B\sim 10^{-9}$G magnetic field
(also see \cite{r24}).

\section{Magnetic field strength and the coherent length in the Universe}

The estimates of a pseudoscalar mass and the photon-pseudoscalar
coupling constant depend essentially on the cosmic magnetic field
in the IGM. There are some of reviews and papers concerning to the
origin and possible effect of magnetic fields in the Universe and
also to the current status of the art of observations of cosmic
magnetic fields (see, for example, \cite{r25}-\cite{r33}).

Typical values of $L_{coh}$ are 100 kpc - 1 Mpc which correspond
to field magnitudes of 10 - 1 $\mu$G. For example, the case of the
Coma Cluster a core magnetic field strength reaches $B\approx 8.3
h^{1/2}_{100} \mu G$ at scales of about 1 kpc. An interesting
example of clusters with a strong magnetic field is the Hydra A
cluster for which the Rotation Measure (RM) implies a 6 $\mu$G
field strength over 100 kpc superimposed with a tangled field of
strength 30 $\mu$G (\cite{r34}). The high-resolution images of
radio sources embedded in galaxy clusters show evidence of strong
magnetic fields in the cluster regions, and also in the regions of
cool fronts and cool fluxes (\cite{r26}). The typical central
field strength is approximately 10 - 30 $\mu$G with the peak value
as large as 10 $\mu$G.

As concerns to the intergalactic medium (IGM), Furlanetto and Loeb
\cite{r29} estimated the magnetic field strength in the diffuse
IGM assuming flux conservation for outflows from QSOs that
inevitably pollute IGM. They obtained $B_0\leq 10^{-9}$G and we
used namely their estimate for calculating constraints on the
coupling constant of photon-pseudoscalar mixing. The same value
has been presented in reviews by Kronberg \cite{r25} and Grasso
and Rubinstein \cite{r35}.

Many authors have considered processes for generating magnetic
fields of cosmological interest. Thus, Langer et al. \cite{r36}
have shown that the photoionization process by photons from the
first cosmic objects provides the magnetic field amplitudes as
high as $2\times 10^{-19}$G. Takahashi et al. \cite{r37} discussed
generation of magnetic field from cosmological perturbations and
showed that the amplitude of produced magnetic field could be
about $\sim 10^{-19}$G at 10 kpc co-moving scale at present.
Siegel and Fry \cite{r38} examined the generation of seed magnetic
fields due to the growth of cosmological perturbations. They
estimated the peak of a magnitude of these fields of $\sim
10^{-30}$G at the epoch of recombination.

As an example one should mention the Biermann mechanism that can
produce seed fields of order $\sim 10^{-19}$G at redshift of
$z\sim 20$ (\cite{r28}).

Rogachevskii et al. \cite{r39} have discussed a new mechanism of
generation of intergalactic large-scale fields in colliding
protogalactic clouds and emerging protostellar clouds.
Self-consistent plasma-neutral gas simulations by Birk et al.
\cite{r40} have shown that seed magnetic field strengths $\leq
10^{-14}$G arise in self-gravitating protogalactic clouds of
spatial scales of 100 pc during $7\times 10^6$ years.

Recently Dolag et al. \cite{r41} have studied the evolution of
magnetic fields in galaxy clusters with use cosmological
magneto-hydrodynamic simulations. They showed that the magnetic
field in core of galaxy clusters for large redshifts $z\sim 3\div
4$ may reach so less magnitude as $B\leq 10^{-14}$G.

Although, there exists essential dispersion of estimates of
intergalactic magnetic field strength, most community prefers the
magnitude of $\sim 10^{-9}$G. Namely, the last data confirm this
point of view. The analysis of the Faraday rotation of the Cosmic
Microwave Background Radiation (CMBR) induced by primordial
magnetic field provides the constraints on its magnitude at the
level $B\sim 10^{-9}$G (\cite{r42}-\cite{r47}).

We use an improved limit on the axion-photon coupling from the
CAST experiment to estimate the magnetic field strength into
intergalactic space. Using the CERN Axion Solar Telescope (CAST),
Andriamonje et al \cite{r23} have set an upper limit on the
axion-photon coupling of $g_{a\gamma}< 8.8\times 10^{-11}
GeV^{-1}$ at 95\% CL from the absence of excess X-rays from the
Sun. This result is now considered as the best experimental limit
over a broad range of axion masses beginning with $m_a < 0.02 eV$.
M$\ddot{o}$rtsell et al. \cite{r24} used this result for the
independent estimate the probability of photon-axion oscillations
in the presence of both intergalactic magnetic fields and an
electron plasma. They found that the current CMB and Csaki et al.
\cite{r2} data can be agreed with the intergalactic magnetic field
strength $B\sim 10^{-9}$G in the framework of the quintessence
model ($\Omega_m = 0.3$, $\Omega_x = 0.7$, $\omega_x = -1/3$) if
the axion mass is $m_a\leq 10^{-10} eV$ and the coupling constant
$g_{a\gamma}\sim 10^{-14} GeV^{-1}$.

We estimate here the intergalactic magnetic field strength at
$z\sim 1$ from the effect of extreme-scale alignments of quasar
polarization vectors revealed by Hutsemekers et al. \cite{r1},
using the CAST upper limit on the axion-like particle-photon
mixing: $g_{a\gamma} < 8.8\times 10^{-11} GeV^{-1}$.

For Eq. (\ref{eq9}) dependence $B_{\bot}(z) = B_0 (1+z)^2$ the
solution of Eqs. (\ref{eq6}) and (\ref{eq7}) gives the next
estimate: $B_0 > 10^{-12} G$.

If we use $B_{\bot}(z) = B_0 (1+z)^{3/2}$ dependence according to
the Chandrasekhar-Fermi law one obtains $B_0 > 4\times 10^{-10}
G$.

Our estimates of the intergalactic magnetic field strength
presented here lie in limits of modern theoretical predictions.

\section{Conclusions}

We estimate the photon-pseudoscalar boson mixing constant from the
effect of cosmological alignment and cosmological rotation of
polarization of distant QSOs revealed by Hutsemekers et al.
\cite{r1}. This effect can be explained in terms of birefringent
phenomenon due to photon-pseudoscalar particle mixing in a
cosmological magnetic field.

We explore two model dependences of cosmological magnetic field on
redshift: quadratic form of $B(z) = B_0 (1+z)^2$ and $B(z) = B_0
(1+z)^{3/2}$. The last relation is based on the
Chandrasekhar-Fermi law.

We obtained the best low constraints on the pseudoscalar-photon
coupling constant: $g_{a\gamma}\leq 4\times 10^{-14} GeV^{-1}$.

On the contrary, one can estimate the strength of the cosmic
magnetic field using the constraints on the axion-like
particle-photon coupling constant from the CAST experiment and
from SNe Ia dimming effect (\cite{r2}). In a result, we obtain the
magnitude of the intergalactic magnetic field ($z\sim 1\div 2$) at
the level of about $4\times 10^{-11}\div 10^{-10} G$.

\section*{Acknowledgements}

We would like to thank for the support by the RFBR (project No.
07-02-00535a), Program of Prezidium of RAS ''Origin and Evolution
of Stars and Galaxies'', the program of the Department of Physical
Sciences of RAS ''Extended Objects in the Universe''.

This research was supported also by the Grant of President of
Russian Federation ''The Basic Scientific Schools''
NS.6110.2008.2.

M.Yu. Piotrovich acknowledges the Council of Grants of President
of Russian Federation for Young Scientists Grant No. 4101.2008.2.

\end{document}